%% file: main.tex
\documentclass[aps,pra,reprint,superscriptaddress]{revtex4-2}

\usepackage{graphicx}
\usepackage{amsmath}
\usepackage{amssymb}
\usepackage{mathtools}
\usepackage{booktabs}
\usepackage{siunitx}
\DeclareSIUnit{\sample}{Sa}
\graphicspath{{figures/}}

\begin{document}

\title{Precision and resource scaling of real-time flux distortion compensation for superconducting quantum control}

\author{Qi Zhou}
\affiliation{CAS Center for Excellence in Quantum Information and Quantum Physics, University of Science and Technology of China, Hefei, Anhui 230026, China}
\affiliation{Laboratory of Quantum Information, University of Science and Technology of China, Hefei, Anhui 230026, China}
\affiliation{Origin Quantum Computing Technology (Hefei) Co., Ltd., Hefei, Anhui 230026, China}
\author{Zi-Hao Mei}
\affiliation{CAS Center for Excellence in Quantum Information and Quantum Physics, University of Science and Technology of China, Hefei, Anhui 230026, China}
\affiliation{Laboratory of Quantum Information, University of Science and Technology of China, Hefei, Anhui 230026, China}
\author{Peng Duan}
\affiliation{Laboratory of Quantum Information, University of Science and Technology of China, Hefei, Anhui 230026, China}
\affiliation{Synergetic Innovation Center of Quantum Information and Quantum Physics, University of Science and Technology of China, Hefei, Anhui 230026, China}
\author{Peng Wang}
\affiliation{CAS Center for Excellence in Quantum Information and Quantum Physics, University of Science and Technology of China, Hefei, Anhui 230026, China}
\affiliation{Laboratory of Quantum Information, University of Science and Technology of China, Hefei, Anhui 230026, China}
\author{Liang-Liang Guo}
\affiliation{Origin Quantum Computing Technology (Hefei) Co., Ltd., Hefei, Anhui 230026, China}
\author{Hao-Ran Tao}
\affiliation{Origin Quantum Computing Technology (Hefei) Co., Ltd., Hefei, Anhui 230026, China}
\author{Wei-Cheng Kong}
\affiliation{Origin Quantum Computing Technology (Hefei) Co., Ltd., Hefei, Anhui 230026, China}
\author{Hui Yang}
\email{yanghui@originqc.com}
\affiliation{Origin Quantum Computing Technology (Hefei) Co., Ltd., Hefei, Anhui 230026, China}
\author{Guo-Ping Guo}
\affiliation{CAS Center for Excellence in Quantum Information and Quantum Physics, University of Science and Technology of China, Hefei, Anhui 230026, China}
\affiliation{Laboratory of Quantum Information, University of Science and Technology of China, Hefei, Anhui 230026, China}
\affiliation{Origin Quantum Computing Technology (Hefei) Co., Ltd., Hefei, Anhui 230026, China}
\affiliation{Institute of Artificial Intelligence, Hefei Comprehensive National Science Center, Hefei, Anhui 230088, China}
\author{Zhao-Yun Chen}
\email{chenzhaoyun@iai.ustc.edu.cn}
\affiliation{Institute of Artificial Intelligence, Hefei Comprehensive National Science Center, Hefei, Anhui 230088, China}


\begin{abstract}
Real-time waveform generation supports dynamic quantum circuits without pre-storing complete waveforms for every execution path. However, long-lived distortions in flux-control lines degrade gate fidelity, requiring compensation to account for the actual pulse history. A frequency-domain inversion and time-domain fitting method is proposed for resource-efficient real-time flux distortion compensation. The method fits the reconstructed compensation impulse response with a compact hybrid infinite impulse response (IIR) and finite impulse response (FIR) filter. Look-ahead parallelization enables this filter to process synthesized waveforms at \SI{1.2}{\giga\sample\per\second} on a field-programmable gate array (FPGA).
Two-qubit cross-entropy benchmarking shows that real-time IIR filtering achieves a median controlled-Z Pauli fidelity close to the software-reference value of $99.57\%$. Numerical analysis and FPGA synthesis indicate approximately logarithmic growth in hardware resource use with compensation timescale. Extending compensation from microsecond to hundred-microsecond timescales increases look-up table (LUT) and digital signal processing (DSP) resource use by only about $14\%$ and $4\%$, respectively, while maintaining a relative arithmetic error below $10^{-4}$. This work provides a scalable hardware foundation for high-fidelity flux control in dynamic superconducting quantum circuits.
\end{abstract}

\maketitle

\input{body}

\begin{acknowledgments}
This work has been supported by the National Key Research and Development Program of China (Grant Nos. 2023YFB4502500 and 2024YFB4504100), the National Natural Science Foundation of China (Grant No. 12404564), and the Anhui Province Science and Technology Innovation (Grant Nos. 202423s06050001 and 202423r06050002).
\end{acknowledgments}

\section*{Data Availability Statement}
The data that support the findings of this study are available from the corresponding author upon reasonable request.

\bibliography{distortion}

\end{document}

%% file: body.tex
\section{Introduction}

Quantum computing has evolved from theoretical exploration to experimental implementation, with research expanding from early noisy intermediate-scale quantum (NISQ) demonstrations toward quantum utility and improved error mitigation~\cite{kim2023evidence,cai2023quantum,robledomoreno2025chemistry}. More recently, tunable superconducting processors have demonstrated quantum error correction below threshold, with logical errors decreasing as the code distance increases~\cite{google2025quantum,lacroix2025scaling}. A universal logical gate set has also been demonstrated in error-detecting surface codes on a superconducting processor~\cite{zhang2025demonstrating}. Extending these advances to fault-tolerant computation places increasing demands on classical feedforward, through which decoded measurement outcomes guide subsequent quantum operations~\cite{caune2026demonstrating}. Dynamic quantum circuits provide this capability by interleaving quantum gates with mid-circuit measurements and outcome-dependent control~\cite{corcoles2021exploiting}. To support such conditional execution, instruction-driven controllers synthesize waveforms from gate-level programs, reducing the need to pre-store complete waveforms for different execution paths~\cite{xu2023qubic,ryan2017hardware}.

\begin{figure*}[!t]
  \centering
  \includegraphics[width=125mm]{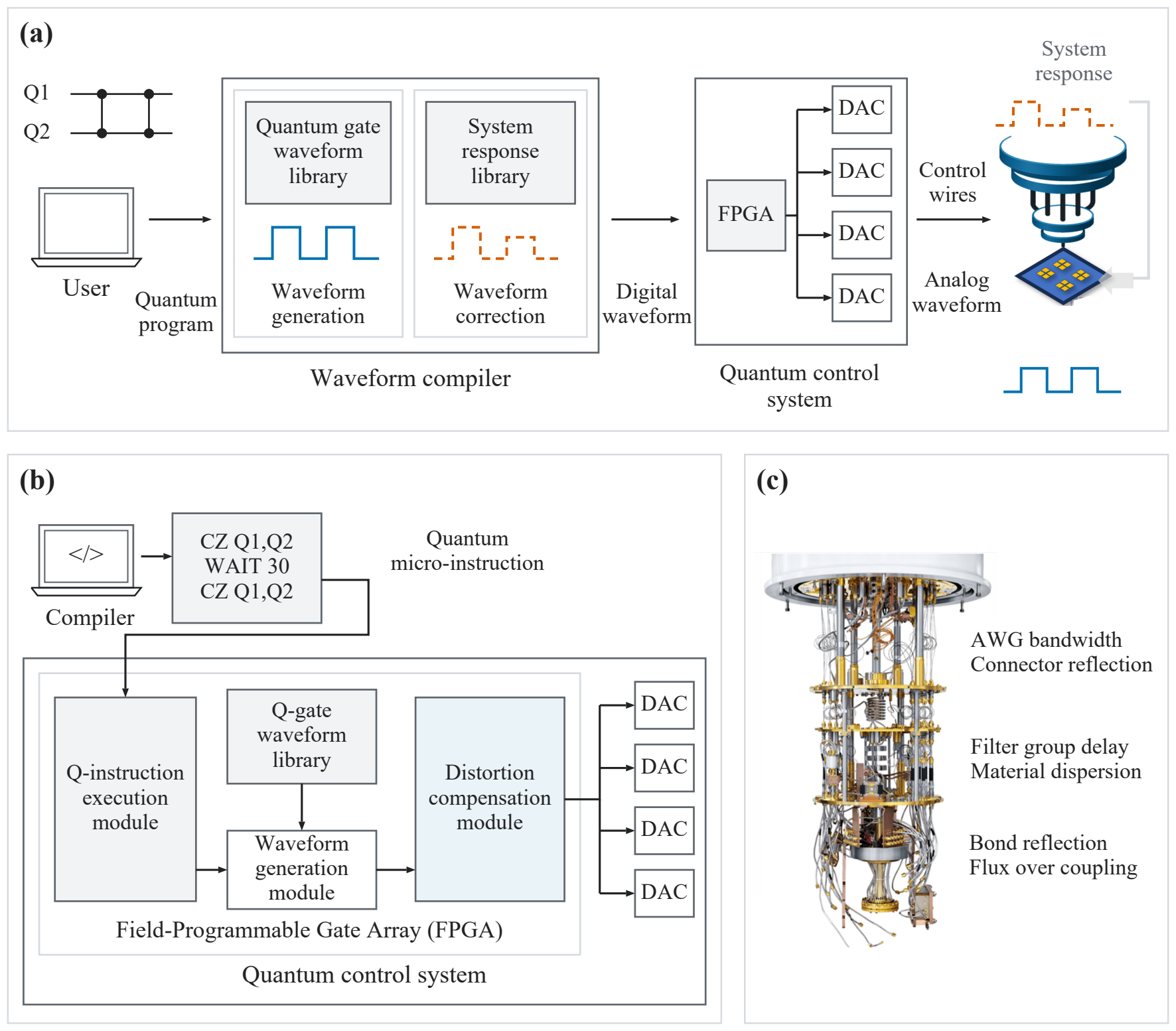}
  \caption{Schematic comparison of distortion-compensation architectures. (a) Offline pre-distortion of complete waveforms before storage and playback. (b) Real-time inverse filtering after gate-level waveform synthesis. (c) Representative flux-line distortion sources in room-temperature electronics (top), the cryogenic transmission chain (middle), and processor interconnects (bottom).}
  \label{fig:intro}
\end{figure*}

However, distortions in the flux-control lines of superconducting processors can persist beyond a single gate, leaving residual responses that affect subsequent pulses~\cite{rol2020time,rol2019fast}. The execution or omission of a conditional flux pulse therefore changes the distortion experienced by later gates. These distortions are commonly addressed using digital pre-distortion based on a characterized flux-line response~\cite{hellings2025calibrating,venkateswaran2026digital,li2025high}. In the conventional offline implementation shown in figure~\ref{fig:intro}(a), the inverse response is applied to the complete waveform before storage and playback. This precomputed correction is tied to a predetermined pulse sequence, whereas dynamic branching changes the pulse history during execution. Real-time inverse filtering after gate-level waveform synthesis, as illustrated in figure~\ref{fig:intro}(b), is therefore essential to track the actual pulse history and maintain consistent control waveforms across dynamic execution paths. A further challenge is to determine how the cost of accurate real-time compensation grows as the flux-line response becomes longer lived.

To this end, we propose a frequency-domain inversion and time-domain fitting (FDI-TDF) method for real-time flux distortion compensation. Building on our previous characterization and calibration framework~\cite{guo2024universal}, the method fits the reconstructed compensation impulse response with a compact hybrid infinite impulse response (IIR) and finite impulse response (FIR) filter to reduce hardware resource requirements. We then design and implement a parallel filtering architecture on a field-programmable gate array (FPGA) using look-ahead transformations. By processing multiple samples per clock, the architecture matches the throughput of gigasample-per-second digital-to-analog converters (DACs) at a lower logic clock rate, sustaining \SI{1.2}{\giga\sample\per\second} after gate-level waveform synthesis.

In addition, we experimentally evaluate the effect of real-time flux distortion compensation on two-qubit gate fidelity. Two-qubit cross-entropy benchmarking (XEB) shows that the fourth-order, 44-bit IIR implementation achieves a median controlled-Z (CZ) Pauli fidelity that closely matches the software-reference median of $99.57\%$. Beyond experimental validation, we investigate how numerical precision and hardware cost scale with compensation timescale at fixed error thresholds, using models derived from qubit-probe measurements of 147 cryogenic flux-control channels. Over timescales from microseconds to hundreds of milliseconds, hardware resource use grows approximately logarithmically. Extending coverage from microsecond to hundred-microsecond timescales increases look-up table (LUT) and digital signal processing (DSP) resource use by approximately $14\%$ and $4\%$, respectively, with a relative arithmetic error below $10^{-4}$. At 44 bits, resource estimates indicate a capacity of nine independent compensation channels per FPGA, matching the scale of eight-output modules used in representative superconducting-qubit control systems~\cite{zhou2026hima,stefanazzi2022qick,singhal2023sqcars}. Together, these results provide a resource-efficient foundation for real-time flux control in future superconducting processors with thousands of physical qubits, supporting progress toward fault-tolerant quantum computing.

\section{Methods}

In this section, we describe the modeling, reconstruction, and parallel implementation required for real-time distortion compensation. We first establish a model of the flux-line response and relate it to time-domain measurements. Building on this model, FDI-TDF constructs a compact compensation filter with reduced hardware resource requirements. Finally, parallel implementation enables the filter to match the waveform sampling rate at the lower FPGA clock frequency.

\subsection{LTI system modeling and time-domain mapping}
\label{subsec:system_modeling}

To construct the inverse filter, we approximate the transfer from digital samples through the coaxial cables, bias-tees, and filters to on-chip flux as a linear time-invariant (LTI) system. For input samples $x[n]$ and impulse response $h[n]$, the on-chip flux waveform is $y[n]=(h*x)[n]$. A rational transfer function describes the amplitude and phase response of the flux line:

\begin{equation}
    H(z)=\frac{B(z)}{A(z)}
    =\frac{\sum_{m=0}^{M_H} b_m^{(H)}z^{-m}}
    {1+\sum_{j=1}^{N_H} a_j^{(H)}z^{-j}}.
    \label{eq:Hz_rational}
\end{equation}
Here $M_H$ and $N_H$ are the numerator and denominator orders of the forward response model.

We use partial fraction decomposition, as in our previous work~\cite{guo2024universal}, to separate the response into components that can be parameterized and implemented individually. For a model with non-repeated poles, this decomposition yields a finite FIR term and first- and second-order IIR modes:

\begin{equation}
    H(z)=\underbrace{H_0(z)}_{\mathrm{FIR}}
    +\underbrace{\sum_{l=1}^{N_1}C_{1,l}H_{1,l}(z)}_{\text{first-order IIR}}
    +\underbrace{\sum_{k=1}^{N_2}H_{2,k}(z)}_{\text{second-order IIR}}.
    \label{eq:Hz_decomp}
\end{equation}
These three components describe the short-time dynamics, monotonic relaxation, and damped oscillations of the flux-line response.

\begin{enumerate}
\item Short-time response: $H_0(z)=\sum_{s=0}^{S}d_sz^{-s}$ represents the prompt response and edge dynamics over a finite sample window. These dynamics reflect the DAC's zero-order-hold response and the bandwidth and propagation characteristics of the transmission chain, including cable skin-effect losses and reflections.

\item Monotonic decay: real-pole sections describe exponential relaxation, as can arise from charging and discharging in bias-tee networks. Each section has the form

\begin{equation}
    H_{1,l}(z)=\frac{1}{1-p_lz^{-1}},\qquad 0<p_l<1.
    \label{eq:first_order_mode}
\end{equation}

\item Damped oscillation: complex-conjugate poles describe resonant components associated with filters and impedance discontinuities in the transmission chain. Pairing the conjugate poles and residues gives a real second-order section,

\begin{equation}
    H_{2,k}(z)=\frac{c_{0,k}+c_{1,k}z^{-1}}
    {1-2r_k\cos\Omega_k\,z^{-1}+r_k^2z^{-2}},\qquad 0<r_k<1.
    \label{eq:second_order_mode}
\end{equation}

The real numerator coefficients $c_{0,k}$ and $c_{1,k}$ set the modal amplitude and phase.
\end{enumerate}

To determine the model parameters experimentally, we use the flux-line step response, which captures both the edge dynamics and the subsequent settling tail~\cite{li2025high,foxen2020demonstrating}. With sampling period $T_s$, the real pole maps to a decay time through $p_l=e^{-T_s/\tau_l}$. A pole approaching the unit circle therefore represents a longer tail; a dominant real mode has the forward-response time constant $\tau=\max_l\tau_l$. For an oscillatory mode, $r_k=e^{-T_s/\tau_k}$ and $\Omega_k=2\pi f_kT_s$ specify its decay time and oscillation frequency. These recursive modes retain long-time response components with a small number of state variables, avoiding a separate FIR coefficient for every sample in the tail.

Using these modal time dependences, we parameterize the measured step response for $n\geq0$ as

\begin{equation}
\begin{split}
y_{\text{model}}[n] &= \sum_{s=0}^{S} C_{0,s} u[n-s] + \sum_{l=1}^{N_1} A_l (1 - e^{-n T_s/\tau_l}) \\
&\quad + \sum_{k=1}^{N_2} B_k \left[1 - e^{-n T_s/\tau_k} \cos(2\pi f_k n T_s + \phi_k)\right].
\end{split}
\label{eq:time_domain_model}
\end{equation}

Here $u[n]$ is the unit-step sequence, $N_1$ and $N_2$ count the monotonic and oscillatory modes, and $S+1$ is the number of FIR taps. The fitting amplitudes and the FIR coefficients absorb the modal gains and initial-value terms of the discrete step response.

We reconstruct the step response using the qubit-based characterization developed in our previous work~\cite{guo2024universal}. The calibration workflow also refines the inverse-filter parameters through a Nelder--Mead search using the mean Clifford fidelity from randomized benchmarking as the objective. The measurements and closed-loop refinement are described in appendix~\ref{app:calibration}.

\subsection{FDI-TDF algorithm and parallel model reconstruction}
\label{subsec:fdi_tdf}

A detailed response model can be inverted numerically in a double-precision software workflow, but its direct hardware realization may require many coefficients and recursive operations. For $H(z)=B(z)/A(z)$, the exact inverse is $A(z)/B(z)$, so its poles are the uncancelled zeros of the forward response. When the forward response is represented by a high-order FIR model, inversion generally yields a high-order IIR filter. High-order IIR filters require more arithmetic operations and stored states, while inverse poles close to the unit circle increase sensitivity to finite-word-length errors.

\begin{figure*}[!t]
  \centering
  \includegraphics[width=125mm]{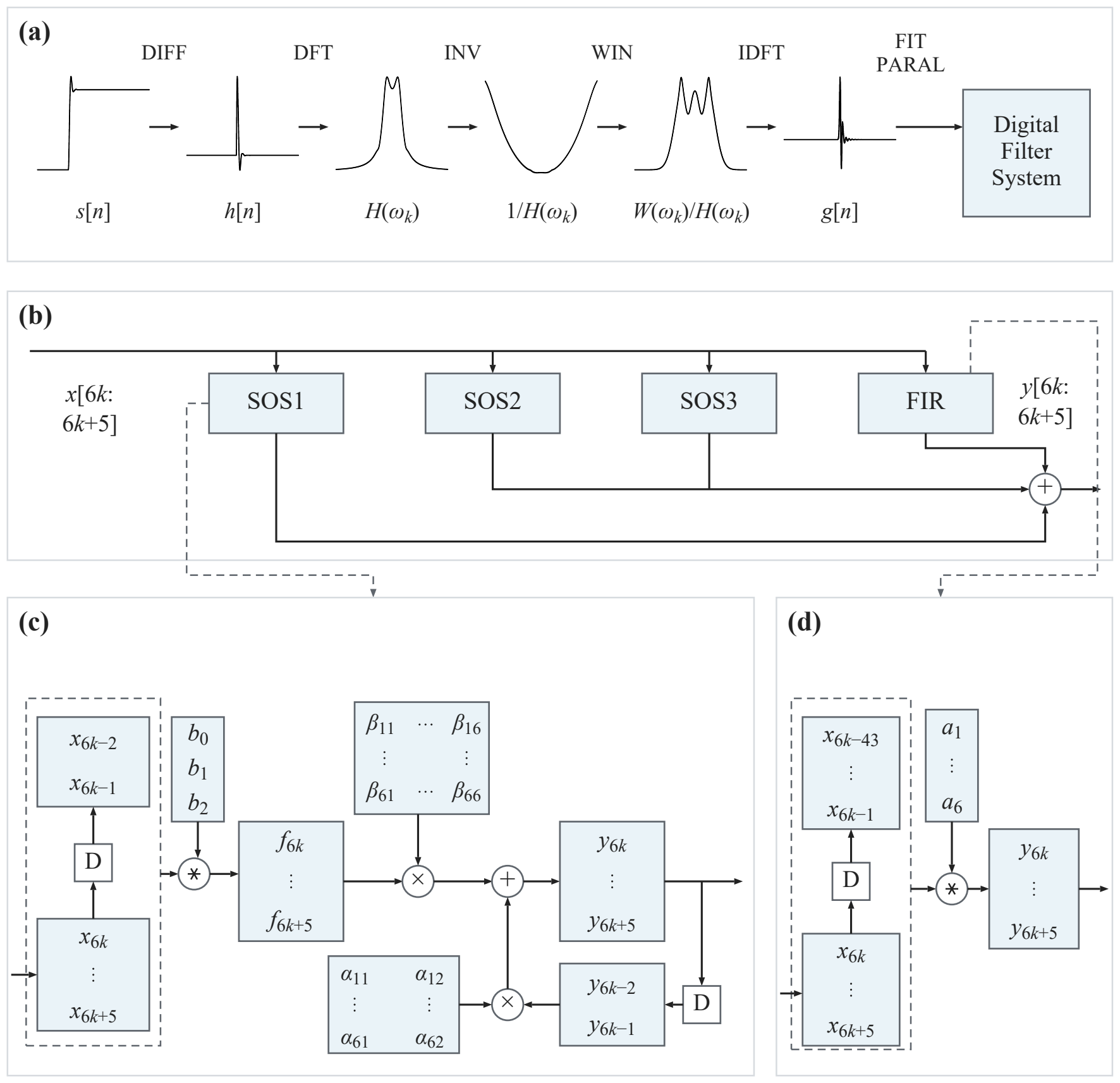}
  \caption{Inverse-filter reconstruction and parallel filtering scheme. (a) FDI-TDF workflow from the measured step response $s[n]$ to the reconstructed compensation impulse response $g[n]$ and compact IIR/FIR model. (b) Three parallel second-order IIR sections (SOS) and a 44-tap FIR branch represent the long- and short-time components of the compensation response. (c) Logical structure of block look-ahead IIR filtering and (d) sliding-window FIR filtering, each producing six consecutive samples per FPGA clock within one channel. In (c), $\alpha_{ij}$ and $\beta_{ij}$ denote the entries of $\mathit{A}$ and $\mathit{B}$ in equation~\eqref{eq:iir_block_methods}, and $D$ denotes a delay of one FPGA clock cycle.}

  \label{fig:method}
\end{figure*}

To address these implementation constraints, we propose frequency-domain inversion and time-domain fitting (FDI-TDF). Frequency-domain inversion reconstructs the target compensation impulse response $g[n]$, and time-domain fitting represents it with a compact FIR branch and low-order recursive sections. The workflow in figure~\ref{fig:method}(a) contains four steps.

\begin{enumerate}
\item Impulse response extraction: we measure the step response $s[n]$ of the flux line and apply the first-order difference $h[n]=s[n]-s[n-1]$ to obtain its discrete impulse response.

\item Frequency-domain inversion with regularization: an $N_{\mathrm{DFT}}$-point discrete Fourier transform (DFT) converts $h[n]$ into the frequency response $H(\omega_k)$. The finite bandwidth of the flux-control chain attenuates its high-frequency response. Direct inversion therefore produces large gains in this region, increasing sensitivity to measurement noise. We reduce the high-frequency contribution with a Gaussian weight $W(\omega_k)=\exp[-(\omega_k/\omega_c)^2/2]$ and define the target inverse response as

\begin{equation}
    G(\omega_k)=\frac{W(\omega_k)}{H(\omega_k)}.
    \label{eq:Gk_weighted}
\end{equation}

\noindent The Gaussian width $\omega_c$ is set to $80\%$--$90\%$ of the Nyquist limit. This smooth low-pass weighting suppresses high-frequency noise amplification while retaining the inverse phase correction and approximately preserving the amplitude correction at frequencies well below $\omega_c$. The relation $H(\omega_k)G(\omega_k)=W(\omega_k)$ specifies a Gaussian-smoothed target waveform for the compensated flux line.

\item Target response reconstruction: an inverse DFT (IDFT) of $G(\omega_k)$ yields the compensation-filter impulse response $g[n]$.

\item Model parameterization by time-domain fitting: we use nonlinear least squares to fit $g[n]$ with the impulse response of a compact hybrid IIR/FIR model. The fitted response contains a finite short-time component and exponential or damped-oscillatory components, whose discrete-time representation defines the compensator $H_{\mathrm{cal}}(z)$.
\end{enumerate}

For FPGA implementation, we express the fitted compensator as a finite causal FIR branch in parallel with real-coefficient second-order IIR sections (SOS):

\begin{equation}
    H_{\text{cal}}(z)=\underbrace{\sum_{m=0}^{M}b_mz^{-m}}_{H_{\text{FIR}}(z)}
    +\underbrace{\sum_{k=1}^{K}\frac{\beta_{0,k}+\beta_{1,k}z^{-1}+\beta_{2,k}z^{-2}}
    {1+\gamma_{1,k}z^{-1}+\gamma_{2,k}z^{-2}}}_{H_{\text{IIR}}(z)}.
    \label{eq:parallel_model}
\end{equation}

Each SOS represents two real poles or a complex-conjugate pair. For stability, the inverse-filter poles $\lambda_{k,j}$ must lie inside the unit circle, $|\lambda_{k,j}|<1$.

As shown in figure~\ref{fig:method}(b), the FIR branch describes short-time correction, while the SOS branches represent the decay and oscillatory components of the inverse response. Each branch processes the same input, and their outputs are summed to produce the compensation waveform. The recursive IIR branches represent long decay tails with a small set of states, without requiring additional FIR taps. The resulting hardware cost therefore depends on the arithmetic precision needed to retain this response history accurately. Our calibration experience indicates that three parallel SOS and a 43rd-order FIR branch are generally sufficient for flux distortion compensation in superconducting processors, with accuracy comparable to software pre-distortion. We therefore adopt this sixth-order IIR structure for the precision and resource evaluation presented in section~\ref{sec:scalability}.

\subsection{Parallel filter architecture via look-ahead transformation and sliding windows}
\label{sec:calib:parallel}

The compact model reduces the number of filter coefficients, but a serial implementation would still require each recursive update to finish at the DAC sampling rate. We address this rate constraint by applying a look-ahead transformation to the IIR sections and a sliding-window construction to the FIR branch. Both branches then compute the same block of consecutive samples in parallel.

\subsubsection{IIR parallelization via look-ahead transformation}

The recursive dependence of $y[n]$ on preceding outputs limits ordinary pipelining of an IIR filter. To address this dependence, we use a look-ahead transformation to express several outputs directly in terms of retained states and known feedforward terms. Consider the serial difference equation for an $N$th-order IIR section:

\begin{equation}
    y[n]=-\sum_{j=1}^{N}a_jy[n-j]+f[n].
    \label{eq:iir_recursion_methods}
\end{equation}
Here $f[n]=\sum_m b_mx[n-m]$ is the feedforward convolution for that section.

For $L$-way parallel processing, we recursively substitute the difference equation until each output $y[n+i]$, with $0\leq i<L$, depends only on outputs preceding the block and on its known feedforward terms. This removes the sequential dependence between outputs within the block and gives

\begin{equation}
    \mathbf{y}_{[n:n+L-1]}=
    \mathit{A}\,\mathbf{y}_{[n-N:n-1]}
    +\mathit{B}\,\mathbf{f}_{[n:n+L-1]}.
    \label{eq:iir_block_methods}
\end{equation}

The retained state is ordered chronologically as $\mathbf{y}_{[n-N:n-1]}=[y[n-N],\ldots,y[n-1]]^{\mathrm T}$, and $\mathbf{f}_{[n:n+L-1]}$ contains the current block of feedforward terms. The coefficient matrices $\mathit{A}\in\mathbb{R}^{L\times N}$ and $\mathit{B}\in\mathbb{R}^{L\times L}$ follow from the recursive expansion; their derivation is given in appendix~\ref{app:iir}.

For second-order sections ($N=2$), figure~\ref{fig:method}(c) illustrates the block computation with $L=6$. At FPGA clock cycle $k$, the two paths contribute to the output block $\mathbf{y}_{[6k:6k+5]}$ as follows.

\begin{enumerate}
\item Feedforward path: the six current input samples and the final two inputs of the preceding block are convolved with the section numerator coefficients $b_0,b_1,b_2$ to form $\mathbf{f}_{[6k:6k+5]}$, which is multiplied by $\mathit{B}$.

\item Feedback path: the previous block supplies the state $[y[6k-2],y[6k-1]]^{\mathrm T}$, which is multiplied by $\mathit{A}$ to propagate the recursive history.
\end{enumerate}

The two contributions are added for each output, and the final two samples, $y[6k+4]$ and $y[6k+5]$, become the state for the next cycle. This block formulation exposes parallel multiplication and addition operations, allowing the filter to match the DAC sampling rate at a lower FPGA clock frequency while retaining the recursive history.

\subsubsection{FIR parallelization via a sliding window}

The FIR branch contains only feedforward operations, so its consecutive outputs can be computed independently from overlapping input windows. An $M$th-order FIR filter uses $M+1$ taps for each output; computing $L$ consecutive outputs therefore requires the input window $x[n-M],\ldots,x[n+L-1]$, containing $M+L$ samples. For the 43rd-order FIR branch and $L=6$ in figure~\ref{fig:method}(d), this gives a 49-sample window for six 44-tap convolutions.

At each clock cycle, the window advances by six samples: six new inputs enter and the six oldest samples leave. Six parallel multiply-accumulate paths use fixed offsets into this window, each evaluating the inner product of the 44 coefficients with its corresponding input sub-window. Together they produce the FIR output block $\mathbf{y}_{[n:n+5]}$ at the same rate as the IIR sections.

Summing the corresponding FIR and IIR outputs yields six consecutive compensated samples per clock within one channel. At a \SI{200}{\mega\hertz} logic clock, the architecture sustains \SI{1.2}{\giga\sample\per\second}. The retained IIR state and FIR input window carry the response history across successive waveform segments, enabling continuous compensation after gate-level waveform synthesis.

\section{Hardware implementation and experimental validation}
\label{sec:hardware_validation}

In this section, we validate the proposed real-time compensation engine through waveform measurements and two-qubit gate benchmarking. We first describe the FPGA hardware platform and compare real-time compensation with double-precision software pre-distortion using room-temperature waveform measurements. In situ qubit measurements then characterize cryogenic flux-line compensation at different IIR word lengths, showing that the 44-bit implementation yields a step response close to the software reference. Building on this waveform validation, two-qubit cross-entropy benchmarking (XEB) evaluates gate performance across different IIR filter orders and word lengths, examining the critical trade-off between arithmetic precision and control performance.

\subsection{Hardware platform and filter architecture}

Experimental validation was performed on a superconducting quantum control system with a Xilinx FPGA-based arbitrary waveform generator (AWG). The flux-control module uses a 16-bit DAC operating at \SI{1.2}{\giga\sample\per\second}, whereas the FPGA fabric is clocked at \SI{200}{\mega\hertz}. To match the DAC sampling rate, the compensation engine processes six consecutive samples per FPGA clock cycle. The primary chip experiments use a fourth-order IIR filter (two second-order sections, SOS) in parallel with a 43rd-order FIR stage (44 taps), with 44-bit arithmetic in both branches.

To benchmark the implemented filter, we used two signal-delivery modes:
\begin{itemize}
    \item \textbf{Real-time compensation:} The FPGA filter continuously processes samples after gate-level waveform synthesis and before DAC conversion.

    \item \textbf{Bypass:} Complete waveforms are computed in software and stored in FPGA memory. Playback bypasses the FPGA filter, providing either uncompensated waveforms or an offline double-precision compensation reference.
\end{itemize}
Both modes use the same DAC channel, cryostat cabling, and qubit flux line.

Room-temperature characterization used a Keysight MSOV334A digital oscilloscope at a sampling rate of \SI{80}{\giga\sample\per\second}. All cryogenic response and two-qubit benchmarking data were acquired on the 102-qubit OriginQ Wukong superconducting quantum processor, which is accessible through the OriginQ cloud platform~\cite{OriginQC_Cloud}.

\subsection{Room-temperature waveform comparison}
\label{subsec:rt_validation}

\begin{figure}[!htb]
  \centering
  \includegraphics[width=\columnwidth]{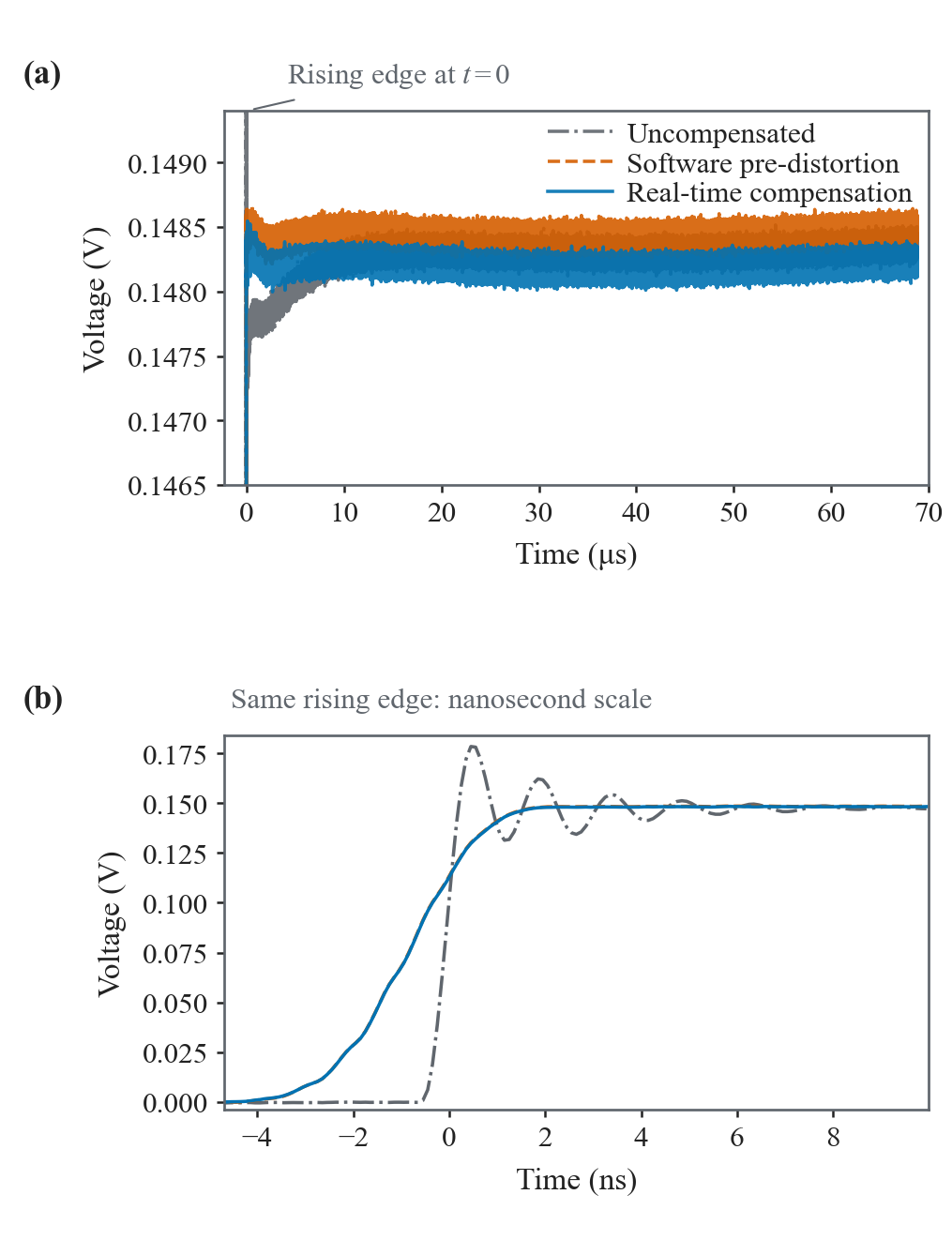}
  \caption{Room-temperature step responses for a \SI{500}{\micro\second}-period square wave: uncompensated (gray dash-dotted), software pre-distorted (orange dashed), and FPGA compensated (blue solid). Each trace averages 1000 waveforms. (a) Microsecond-scale settling and (b) a nanosecond-scale view of the same rising edge. The FPGA configuration uses a fourth-order IIR branch and an FIR branch with 14 effective taps.}
  \label{fig:rt_comparison}
\end{figure}

We first compare software pre-distortion and real-time compensation at room temperature using the fourth-order IIR filter and FIR branch described above. For this experiment, an effective FIR order of 13 is sufficient within the implemented 43rd-order FIR structure.

To measure the step response, the AWG generated a square wave with a period of \SI{500}{\micro\second}. Because the high-level duration substantially exceeds the dominant response time constant, each rising edge provides a close approximation to an isolated step. The output was recorded with the oscilloscope and averaged over 1000 waveforms to reduce measurement noise.

Figure~\ref{fig:rt_comparison}(a) shows that both compensation modes suppress the microsecond-scale settling tail of the uncompensated response. The FPGA output also reproduces the edge shaping and ringing suppression of the software reference, as seen in the nanosecond-scale view of the same rising edge in figure~\ref{fig:rt_comparison}(b). The FPGA-compensated plateau voltage is approximately 0.2\% lower than the software-compensated value. This static offset can be absorbed into the routine calibration of the qubit operating point, without requiring an additional compensation step. The similar settling and edge responses therefore demonstrate that the real-time engine reproduces the principal transient correction of software pre-distortion.

\subsection{Cryogenic distortion calibration}
\label{subsec:cryo_validation}

We next use in situ qubit measurements to assess how arithmetic precision affects cryogenic flux-line compensation, comparing fourth-order IIR filters at 35 and 44 bits with software pre-distortion. Because the line impedance and filtering characteristics depend on temperature, we reconstruct the step response under cryogenic operating conditions and determine the compensation filter using the procedure described in appendix~\ref{app:calibration}. The measured flux-line response has a dominant time constant of approximately \SI{4.8}{\micro\second}.

\begin{figure*}[t]
  \centering
  \includegraphics[width=125mm]{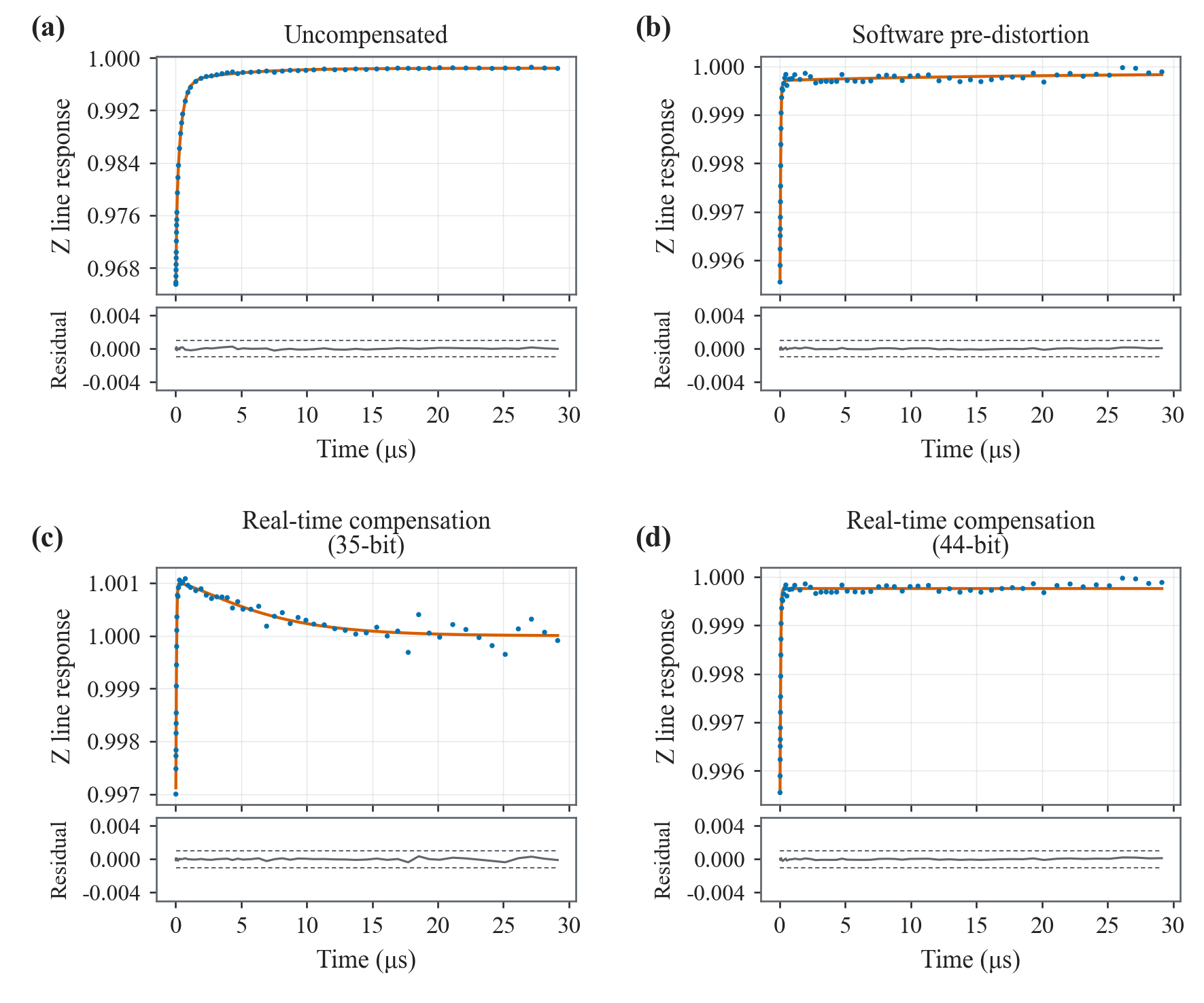}
  \caption{Cryogenic flux-line step responses measured with the qubit. Blue points are measurements and orange curves are fits; gray traces beneath each response show the fit residuals, with dashed reference lines at $\pm10^{-3}$. (a) Uncompensated; (b) software pre-distorted; (c), (d) FPGA compensated using fourth-order IIR filters at 35 and 44 bits, respectively. The 44-tap FIR branch uses 44-bit arithmetic.}
  \label{fig:cryo_distort}
\end{figure*}

Figure~\ref{fig:cryo_distort}(a) shows the pronounced settling tail of the uncompensated response. The 35-bit implementation in figure~\ref{fig:cryo_distort}(c) suppresses the dominant distortion; however, its normalized response decreases from approximately 1.001 to 1.000, leaving a tail with an initial amplitude of about $10^{-3}$ ($0.1\%$) whose decay extends beyond \SI{10}{\micro\second}. In contrast, the 44-bit response in figure~\ref{fig:cryo_distort}(d) is nearly flat and approaches the software reference in figure~\ref{fig:cryo_distort}(b), with measured fluctuations on the order of $10^{-4}$. The gray traces beneath each response show deviations of the measurements from their fitted responses, also on the order of $10^{-4}$.

This improvement is consistent with the reduction of coefficient quantization and state rounding errors when wider arithmetic is used in the recursive filter. Thus, the 44-bit implementation yields cryogenic waveform compensation close to the software reference.

\subsection{Two-qubit gate benchmarking}
\label{subsec:gate_benchmarking}

\begin{figure*}[t]
  \centering
  \includegraphics[width=125mm]{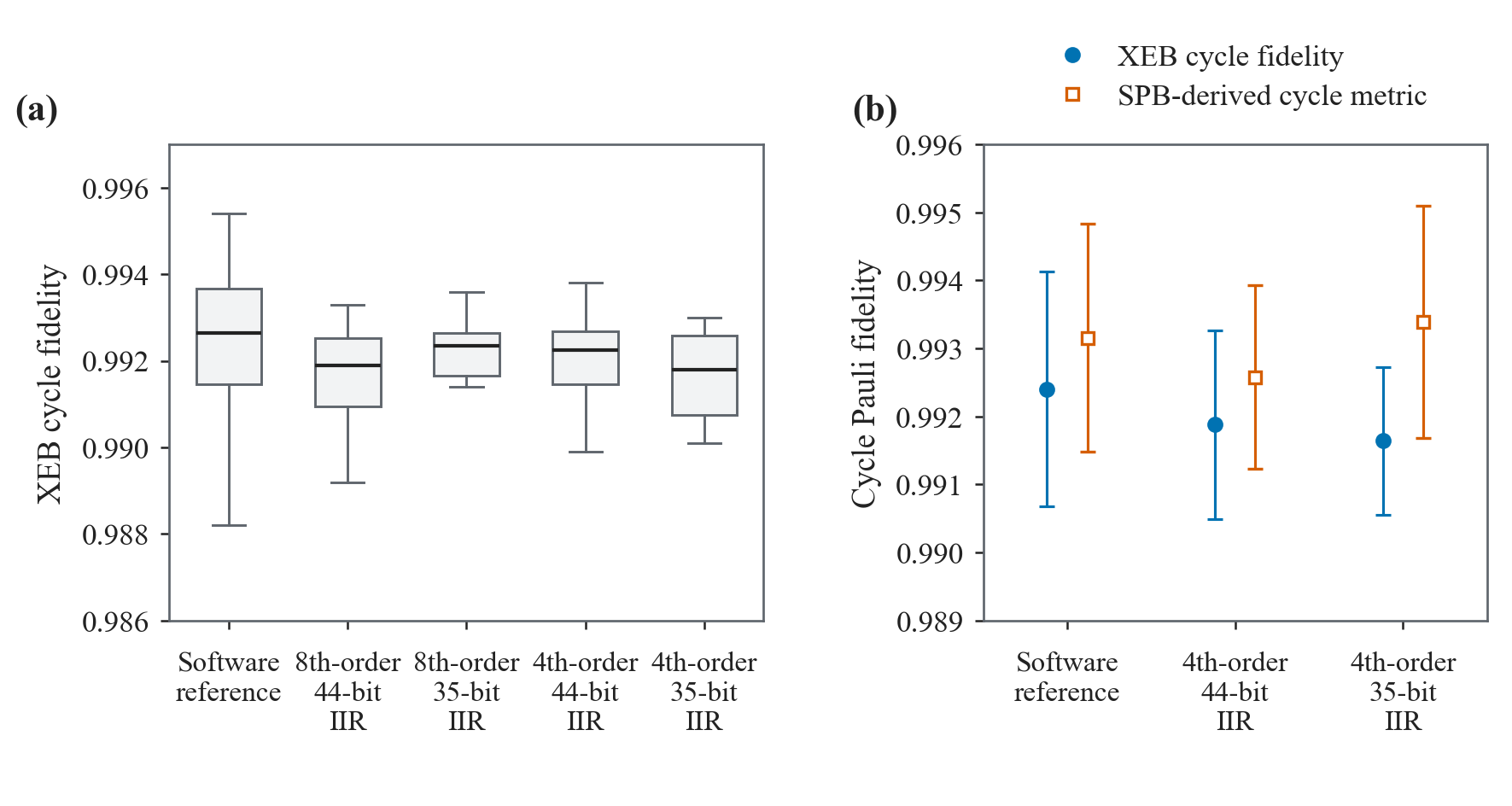}
  \caption{Benchmarking results in the cycle Pauli fidelity convention. (a) XEB distributions for the software reference (40 rounds) and four IIR configurations (10 rounds each), labeled by order and word length. Boxes span the interquartile range; black lines mark medians. (b) Mean XEB fidelity (blue circles) and SPB-derived cycle metric (orange open squares) for the software reference and fourth-order configurations. Error bars are one sample standard deviation.}
  \label{fig:fidelity_limit_comparison}
\end{figure*}

To extend the validation from waveform accuracy to gate performance, we performed two-qubit cross-entropy benchmarking (XEB) across different IIR filter orders and word lengths. The measurements used a qubit pair ($q_1, q_2$) from the OriginQ Wukong processor, with physical parameters summarized in table~\ref{tab:qubit_params}.

\begin{table}[!htb]
\centering
\caption{Representative physical parameters of the qubit pair used for cryogenic compensation and benchmarking.}
\label{tab:qubit_params}
\begin{tabular}{lcccc}
\toprule
\textbf{Parameter} & \textbf{Symbol} & \textbf{$q_1$} & \textbf{$q_2$} & \textbf{Unit} \\
\midrule
Qubit frequency       & $f_q$                    & 5.3393  & 5.2056  & \si{\giga\hertz}  \\
Anharmonicity         & $\alpha$                 & $-220.0$ & $-220.3$ & \si{\mega\hertz}  \\
Energy relaxation time & $T_1$                    & 25.3    & 48.5    & \si{\micro\second} \\
Sweet-spot voltage    & $V_{\mathrm{sweet}}$     & $-409$  & $-470$  & \si{\milli\volt}   \\
\bottomrule
\end{tabular}
\end{table}

The single-qubit gate duration is \SI{50}{\nano\second}, with calibrated single-qubit gate errors of $0.001$--$0.002$ for both qubits. The effective coupling rate is \SI{20}{\mega\hertz}, the residual ZZ coupling rate is \SI{0.1}{\kilo\hertz}, and the CZ gate duration is \SI{35}{\nano\second}.

We used the fourth-order, 44-bit IIR filter as the primary configuration for comparison with the double-precision software reference. To examine the effects of order and word length, we also tested a fourth-order, 35-bit filter and eighth-order filters (four SOS) at 35 and 44 bits. Increasing the order allows more response modes to be represented, whereas increasing the word length reduces the arithmetic error in their implementation. The coefficient and state word lengths were both set to the specified IIR width; the 44-tap FIR branch retained 44-bit arithmetic for all configurations.

Each XEB cycle contains one CZ gate and a single-qubit gate on each qubit. We estimate the CZ Pauli fidelity by subtracting a representative single-qubit Pauli error of $0.15\%$ per gate from the cycle error~\cite{arute2019quantum}.

Each circuit was measured using 1000 shots. Environmental fluctuations and drift of the chip parameters can affect the fidelity measured during long XEB acquisitions. Each firmware configuration was therefore evaluated over 10 rounds, requiring approximately \SI{10}{\minute}, to limit the influence of slow drift. The software reference was measured over 40 rounds to characterize the range of fidelity fluctuations under noise and drift and provide a baseline for interpreting the firmware results.

The primary fourth-order, 44-bit configuration gives a median CZ Pauli fidelity estimate of $99.53\%$, close to the software-reference median of $99.57\%$. Figure~\ref{fig:fidelity_limit_comparison}(a) also shows that the XEB medians of all four firmware configurations lie within the software-reference interquartile range. The corresponding CZ Pauli fidelity estimates span $99.48\%$--$99.54\%$. Across these configurations, increasing order or word length does not produce a monotonic improvement in the measured median fidelity.

To assess the contribution of incoherent errors, we extracted the speckle purity benchmarking (SPB) metric from the same XEB records~\cite{arute2019quantum}. This metric is $F_{\mathrm{purity}}=1-e_{\mathrm{P,incoh}}$, where $e_{\mathrm{P,incoh}}$ denotes the inferred incoherent Pauli error, and provides an upper bound on cycle Pauli fidelity under the standard unitarity-benchmarking model~\cite{wallman2015estimating}. Figure~\ref{fig:fidelity_limit_comparison}(b) compares the mean XEB cycle fidelity with the SPB-derived metric for the software reference and the fourth-order configurations. Their small separation of $0.07$--$0.18$ percentage points indicates that incoherent errors account for most of the measured cycle error. Together with the absence of systematic fidelity improvement at higher filter order or word length, these results suggest that residual flux distortion is not the dominant limitation on the measured gate fidelity.

Taken together with the cryogenic calibration results in section~\ref{subsec:cryo_validation}, these measurements show that the fourth-order, 44-bit real-time filter delivers gate performance close to the software reference. The fourth-order, 35-bit case retains a decaying amplitude variation in figure~\ref{fig:cryo_distort}(c), yet its XEB median also lies within the software-reference interquartile range. Thus, improving the compensated step response does not necessarily yield a corresponding improvement in measured gate fidelity. We next consider the implementation requirements for extending this approach to multichannel control.

\section{Numerical precision and resource scaling}
\label{sec:scalability}

In this section, we analyze how arithmetic precision determines compensation accuracy, the supported response timescales, and FPGA resource use. Variations in cabling, cryogenic filtering, and interconnects lead to different settling responses across channels. Longer inverse-filter time constants place the dominant poles closer to the unit circle, increasing sensitivity to coefficient quantization and state rounding~\cite{oppenheim1972effects}. However, wider arithmetic also increases the hardware cost per channel. We therefore first assess the joint influence of coefficient and state precision, then use a common word length to relate time-constant coverage to hardware cost.

For this analysis, we use cryogenic response measurements from $N_{\mathrm{ch}}=147$ selected flux-control channels on the OriginQ Wukong processor, obtained using the qubits as in situ probes. Most of the resulting compensation models can be represented by two SOS, while a small minority require a third section to capture finer response features. We therefore use a sixth-order IIR architecture with six samples processed per clock. The additional SOS provides fitting margin for more complex responses.

\subsection{Impact of arithmetic precision on compensation outputs}
\label{sec:bit_width}

We independently vary the coefficient and state word lengths from 31 to 44 bits and compare the fixed-point outputs with a double-precision software reference.

\begin{figure}[!htb]
	\centering
\includegraphics[width=\columnwidth]{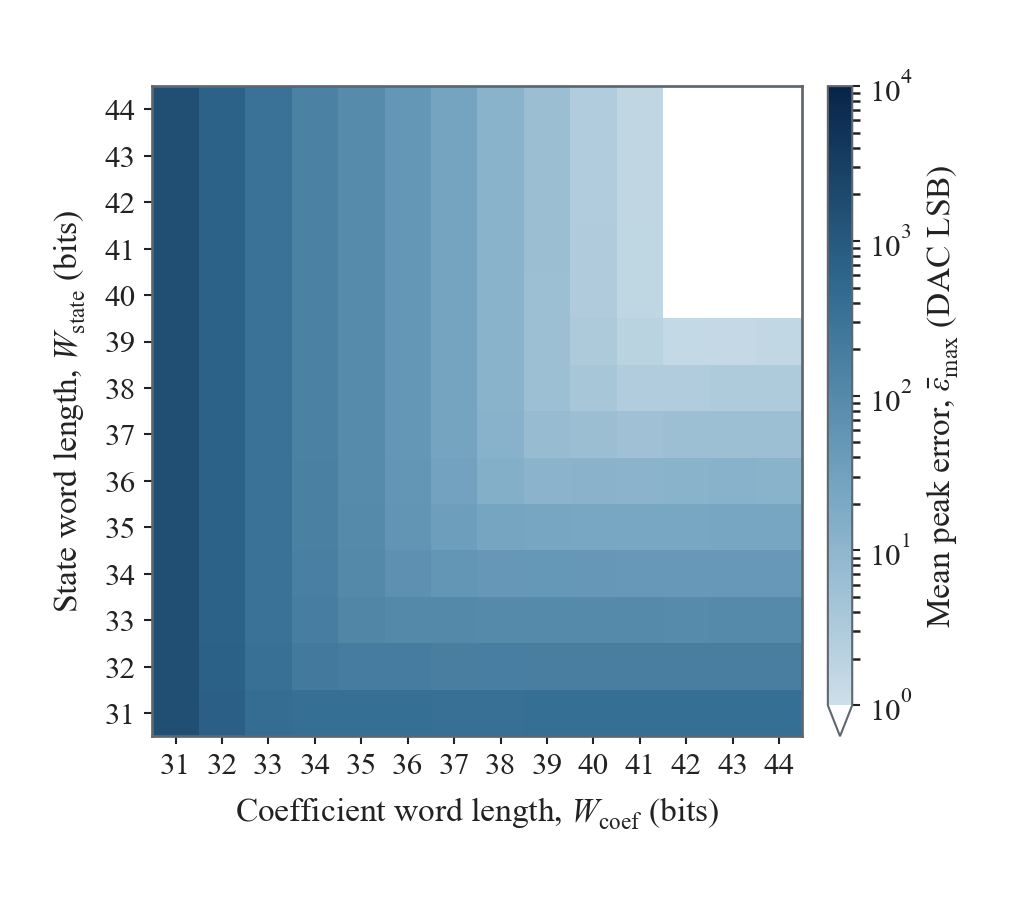}
\caption{Fixed-point error for the compensation models of 147 selected flux-control channels versus coefficient and state word lengths. The maximum absolute difference from the double-precision output is computed for each channel, then averaged across all 147 channels to give $\bar{\epsilon}_{\max}$. The logarithmic color scale uses native DAC-code least significant bits (LSBs); white cells indicate $\bar{\epsilon}_{\max}<1$ LSB.}
\label{fig:bitwidth_accuracy}
\end{figure}

At their native parameters, the compensation models have a mean dominant inverse-pole time constant of \SI{10.44}{\micro\second}. For each $(W_{\mathrm{coef}},W_{\mathrm{state}})$ pair, we apply the same step input to the fixed-point realization and a double-precision implementation of the same IIR model and compare their outputs over 20000 samples. For channel $c$, let $e_c[n]$ denote the sample-wise difference over $N_c$ samples. The channel-averaged maximum absolute error is defined as

\begin{equation}
\bar{\epsilon}_{\max}
=\frac{1}{N_{\mathrm{ch}}}\sum_{c=1}^{N_{\mathrm{ch}}}
\left(\max_{1\leq n\leq N_c}\lvert e_c[n]\rvert\right).
\label{eq:error_metric}
\end{equation}

The numerical error is expressed in units of the 16-bit DAC's least significant bit (LSB), which corresponds to the voltage step between adjacent DAC codes.

Figure~\ref{fig:bitwidth_accuracy} shows an overall decrease in error toward the upper-right region as coefficient and state precision increase together. For the native response models, the channel-averaged peak error is below one LSB when both word lengths are at least 42 bits. The overall diagonal pattern of the error map supports using comparable coefficient and state word lengths. We therefore use equal coefficient and state word lengths in the FPGA implementation to simplify the datapath. In the following analysis, we vary this common word length to evaluate the supported response timescales and hardware resource requirements.

\subsection{Time-constant coverage versus word length}
\label{sec:tau_scaling}

To determine how much precision is needed for longer responses, we varied the dominant inverse-filter time constant $\tau$ in a model family constructed from these compensation models. The sampling rate was $f_s=\SI{1.2}{\giga\sample\per\second}$, and the coefficient and state word lengths were set to a common value, $W_{\mathrm{coef}}=W_{\mathrm{state}}=W$. For each model, all poles were shifted radially by a common factor so that the largest pole magnitude became $\exp[-1/(f_s\tau)]$. The pole angles, relative radii, step-response modal amplitudes, and branch DC gains were retained.

\begin{figure}[!htb]
    \centering
    \includegraphics[width=\columnwidth]{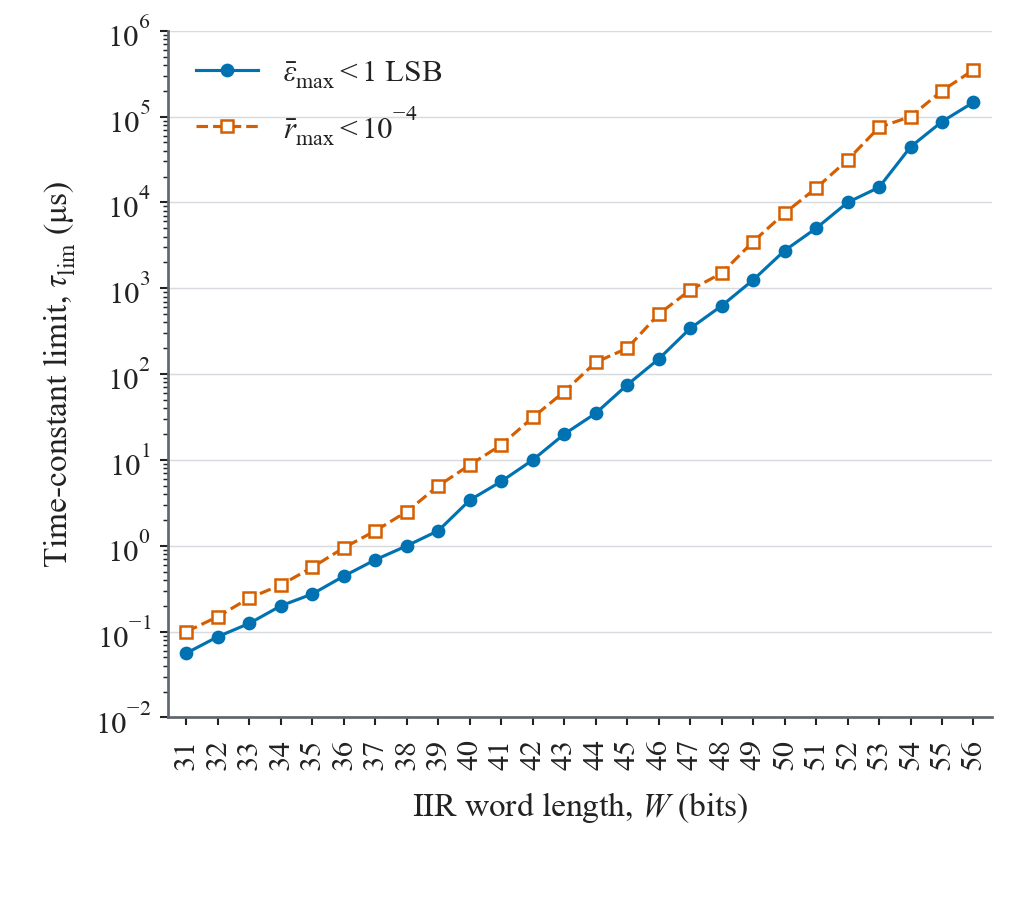}
    \caption{Time-constant coverage of the three-SOS, six-sample-parallel IIR architecture for the 147-model family at \SI{1.2}{\giga\sample\per\second}, with $W_{\mathrm{coef}}=W_{\mathrm{state}}=W$. Blue circles with a solid line show the channel-averaged peak-error criterion $\bar{\epsilon}_{\max}<1$ LSB. Orange open squares with a dashed line show the relative error threshold $\bar{r}_{\max}<10^{-4}$, with errors normalized to each channel's maximum software-reference output amplitude as defined in equation~\eqref{eq:relative_error_metric}.}
    \label{fig:timeconstant_bitwidth}
\end{figure}

We compared the fixed-point output with a high-precision software evaluation of the same IIR model over records spanning at least $8\tau$ and containing at least 20,000 samples. Accuracy is assessed in both absolute and relative terms. The absolute criterion $\bar{\epsilon}_{\max}<1$ LSB compares the arithmetic error in equation~\eqref{eq:error_metric} against the DAC resolution. The relative error is obtained by normalizing each channel's maximum absolute error to the maximum magnitude of its software-reference output $y_{\mathrm{ref},c}[n]$ and then averaging these ratios across channels:

\begin{equation}
\bar{r}_{\max}
=\frac{1}{N_{\mathrm{ch}}}\sum_{c=1}^{N_{\mathrm{ch}}}
\frac{\max_{1\leq n\leq N_c}\lvert e_c[n]\rvert}
{\max_{1\leq n\leq N_c}\lvert y_{\mathrm{ref},c}[n]\rvert}.
\label{eq:relative_error_metric}
\end{equation}

The qubit-probe measurements in figure~\ref{fig:cryo_distort}(b) and (d) show nearly identical responses for software and 44-bit real-time compensation, with relative fluctuations on the order of $10^{-4}$. We therefore require the relative error to satisfy $\bar{r}_{\max}<10^{-4}$. For each word length, we varied $\tau$ to determine the range satisfying each error threshold. Figure~\ref{fig:timeconstant_bitwidth} shows the resulting time-constant coverage $\tau_{\mathrm{lim}}$ for 31--56 bits.

The numerical results show that the required word length increases approximately logarithmically with the supported time constant. At 44 bits, the filter supports inverse-filter time constants up to \SI{35}{\micro\second} under the one-LSB criterion. This range covers the typical response timescales of cryogenic flux-control chains in superconducting quantum processors. Increasing the word length to 56 bits extends the coverage beyond \SI{100}{\milli\second}. With a relative error below $10^{-4}$, the supported time constants extend beyond \SI{100}{\micro\second} at 44 bits and reach several hundred milliseconds at 56 bits.

\subsection{Resource scaling with compensation timescale}
\label{sec:utilization}

Higher arithmetic precision extends the supported response timescales but also increases the hardware resources required per channel. To quantify this dependence, we pair the time-constant coverage in figure~\ref{fig:timeconstant_bitwidth} with the resource utilization in table~\ref{tab:resource} at each common word length.

\begin{table}[!htb]
	\centering
	\caption{Post-synthesis FPGA resource use for one six-sample-parallel IIR engine with three SOS, where $W_{\mathrm{coef}}=W_{\mathrm{state}}=W$. Absolute counts are followed in parentheses by percentages of the corresponding total resources of a Xilinx Zynq UltraScale+ XCZU49DR device~\cite{amd2026ultrascale}.}
	\label{tab:resource}
	\begin{tabular}{crr}
		\toprule
		IIR word length $W$ (bits) & \multicolumn{1}{c}{DSPs (\%)} & \multicolumn{1}{c}{LUTs (\%)}\\
		\midrule
		24 & 180 (4.2\%) & 9,009 (2.1\%) \\
		28 & 378 (8.8\%) & 10,968 (2.6\%) \\
		32 & 378 (8.8\%) & 12,204 (2.9\%) \\
		36 & 429 (10.0\%) & 14,043 (3.3\%) \\
		40 & 446 (10.4\%) & 15,015 (3.5\%) \\
		44 & 446 (10.4\%) & 15,981 (3.8\%) \\
		48 & 770 (18.0\%) & 17,790 (4.2\%) \\
		52 & 770 (18.0\%) & 18,816 (4.4\%) \\
		56 & 821 (19.2\%) & 20,685 (4.9\%) \\
		\bottomrule
	\end{tabular}
\end{table}

\begin{figure*}[!t]
    \centering
    \includegraphics[width=150mm]{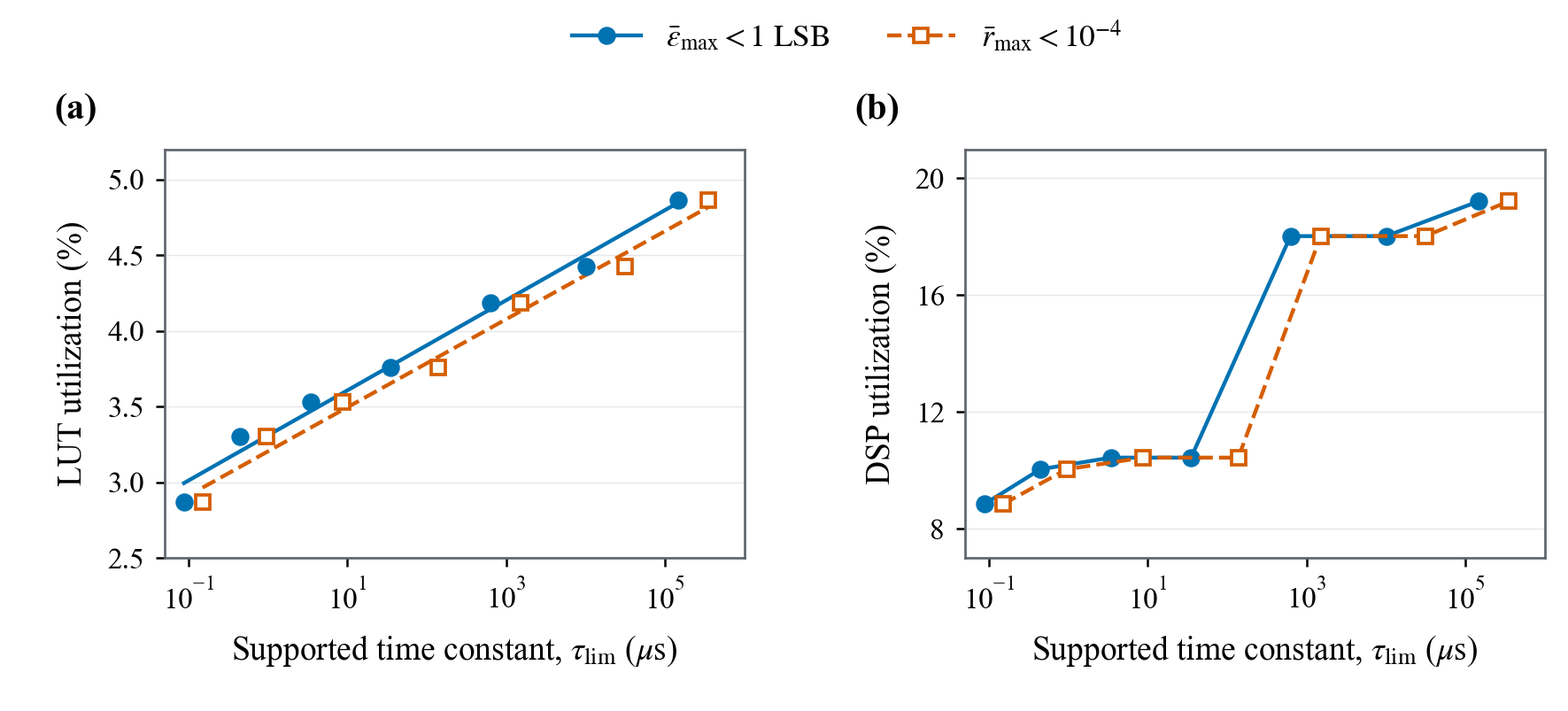}
    \caption{FPGA resource utilization versus supported inverse-filter time constant for the three-SOS, six-sample-parallel IIR architecture at \SI{1.2}{\giga\sample\per\second}. Utilization is expressed as a percentage of the corresponding total resource on the XCZU49DR FPGA. Points pair the unrounded coverage limits in figure~\ref{fig:timeconstant_bitwidth} with the synthesis results in table~\ref{tab:resource} at 32, 36, 40, 44, 48, 52, and 56 bits. Blue circles denote $\bar{\epsilon}_{\max}<1$ LSB; orange open squares denote $\bar{r}_{\max}<10^{-4}$. (a) LUT utilization, with logarithmic fits shown as lines. (b) DSP utilization, with lines connecting the data points.}
    \label{fig:resource_scaling}
\end{figure*}

Figure~\ref{fig:resource_scaling} shows an overall approximately logarithmic increase in LUT and DSP utilization as the supported compensation timescale extends from microseconds to hundreds of milliseconds, with DSP use progressing in discrete steps along this trend. Within the microsecond-to-hundred-microsecond range relevant to superconducting flux control, extending coverage from approximately \SI{1}{\micro\second} to \SI{138}{\micro\second} increases LUT and DSP use by only about $14\%$ and $4\%$, respectively, while maintaining a relative arithmetic error below $10^{-4}$. At 44 bits, resource estimates indicate a capacity for real-time compensation of nine independent channels per FPGA. Higher arithmetic precision extends compensation to hundred-millisecond timescales at the cost of reduced channel parallelism. Furthermore, assigning the IIR order according to each channel's response complexity could improve resource use and multichannel scalability.

\input{discussion}

\appendix

\section{Step-response characterization of the full flux-control chain}
\label{app:calibration}

\begin{figure*}[!t]
    \centering
    \includegraphics[width=125mm]{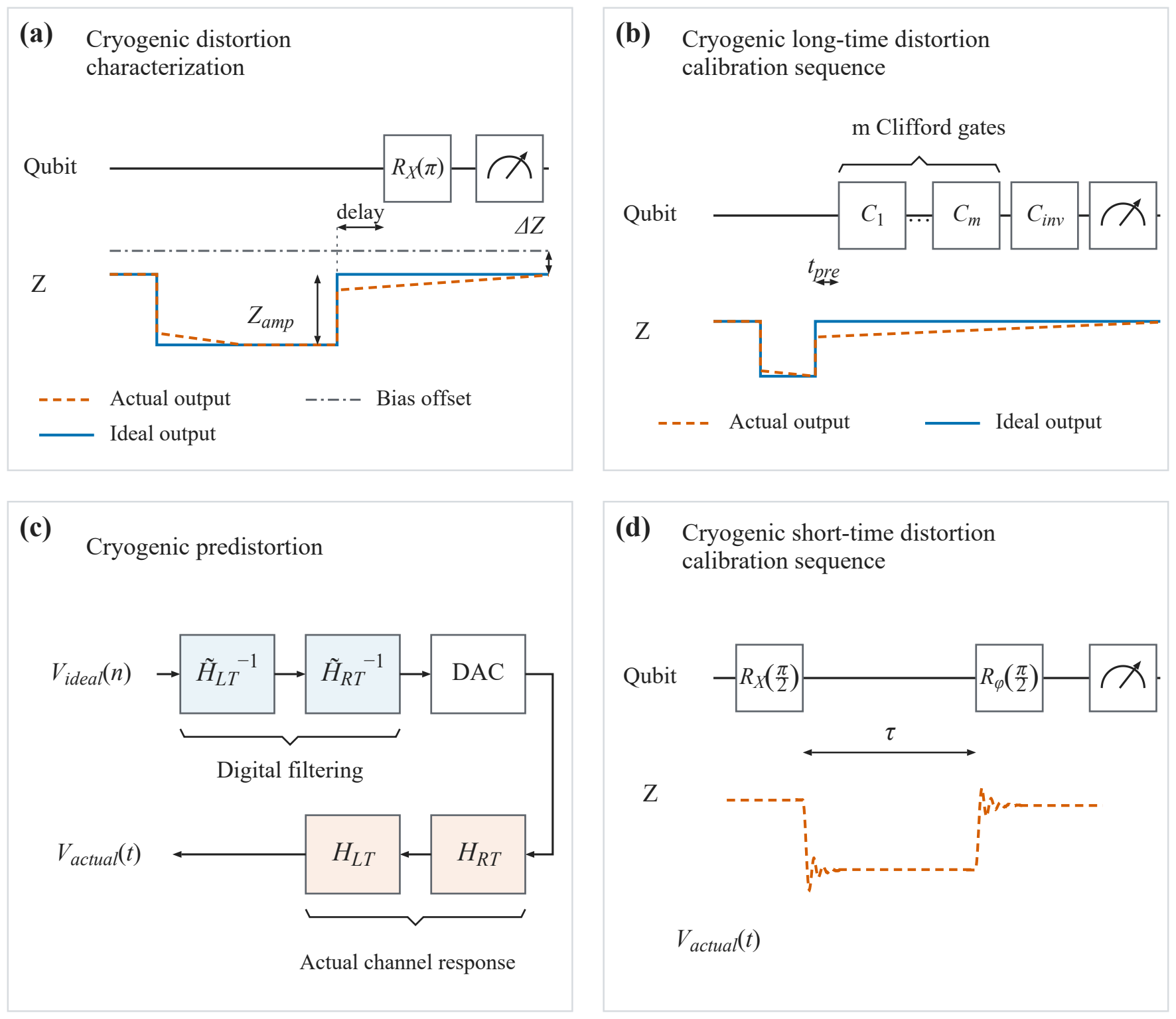}
    \caption{Flux-line calibration workflow. (a) Long-time step-response characterization by sweeping the probe delay and compensation bias $\Delta Z$. (b) Randomized benchmarking following a pre-distorted flux pulse to refine the inverse filter. (c) Room-temperature and cryogenic long-time pre-distortion stages. (d) Cryoscope measurement of the remaining short-time response from the accumulated Ramsey phase and calibrated flux--frequency relation.}
    \label{fig:calib_circuit}
\end{figure*}

To construct an inverse filter for real-time flux compensation, we characterize the \emph{full-chain} step response from the digital waveform samples to the on-chip flux bias, following the approach of reference~\cite{guo2024universal}. We separate the response into (i) a room-temperature electronic part $H_{\mathrm{RT}}$, (ii) a cryogenic long-time part $H_{\mathrm{LT}}$, and (iii) a cryogenic short-time transient. Figure~\ref{fig:calib_circuit}(a)--(d) summarizes the workflow: instrument measurements identify $H_{\mathrm{RT}}$; an in situ qubit measurement initializes the long-time model; an objective based on randomized benchmarking (RB)~\cite{magesan2011scalable} refines $H_{\mathrm{LT}}^{-1}$; and a pre-distortion-assisted Cryoscope measurement resolves the remaining short-time response. Together, these measurements provide the response information needed to reconstruct the compensating filter.

\subsection{Room-temperature electronic characterization}
We drive the room-temperature flux-control output with a long-period square wave and record it with a high-bandwidth oscilloscope. The half-period is much longer than the longest response time constant, so each edge approximates an isolated step, while waveform averaging reduces measurement noise. Fitting the recorded trace to the model in equation~\eqref{eq:time_domain_model} determines the room-temperature response $H_{\mathrm{RT}}(z)$, from which the inverse $H_{\mathrm{RT}}^{-1}(z)$ is constructed. This inverse filter preprocesses subsequent cryogenic probe waveforms, so the residual response is dominated by the cryogenic part of the flux-control chain.

\subsection{Characterization of cryogenic step response}
\label{subsec:step_response_char}

Cryogenic flux lines can exhibit long settling tails~\cite{blais2021circuit,foxen2020demonstrating}. After a control signal returns from $Z_{\mathrm{amp}}$ to the idle bias, the residual flux $b(t)$ decays over a characteristic time $\tau$ and shifts the qubit frequency by $\Delta\omega_q(t)$. The sequence in figure~\ref{fig:calib_circuit}(a) uses the qubit as an in situ flux-sensitive probe of this response.

We apply a periodic square wave of amplitude $Z_{\mathrm{amp}}$ to the flux line, with pulse and idle intervals much longer than the longest response time constant. After the pulse returns to the idle bias, a calibrated $R_x(\pi)$ pulse probes the excited-state population $P_1$ at a variable delay $t$. Residual flux detunes the qubit from the calibrated drive frequency and reduces $P_1$. We therefore sweep an additional bias $\Delta Z$ during the delay and probe window and identify the value $\Delta Z^\star(t)$ that maximizes $P_1$.

Repeating the measurement at delays $\{t_k\}$ gives $b(t_k)\approx-\Delta Z^\star(t_k)$ in the calibrated bias coordinate. Nonlinear least-squares fitting of this reconstructed tail to the response model in equation~\eqref{eq:time_domain_model} yields modal amplitudes and decay times, together with oscillation frequencies and phases when damped oscillations are present. We use these estimates to construct the initial long-time inverse model for closed-loop refinement.

\subsection{Closed-loop refinement of the cryogenic long-time inverse response using randomized benchmarking}
\label{subsec:supp_rb_longtime}

Building on the open-loop response estimate, we refine the long-time inverse filter through the closed-loop RB optimization in figure~\ref{fig:calib_circuit}(b). Residual flux after a large Z step produces a detuning $\Delta\omega_q(t)$ and an accumulated phase error $\delta\phi\approx\int\Delta\omega_q(t)\,\mathrm{d}t$. Because this error affects subsequent gates, we use the mean Clifford fidelity as the objective for optimizing the inverse-filter parameters~\cite{kelly2014optimal}.

The experimental execution of this closed-loop calibration involves four key stages:

\begin{enumerate}
    \item \textbf{Pre-distorted excitation:} A rectangular pulse of amplitude $Z_{\mathrm{amp}}$ excites the long-time response and is compensated using the current candidate filter $\tilde{H}^{-1}_{\mathrm{LT}}(z;\boldsymbol{\theta})$, where $\boldsymbol{\theta}$ contains the adjustable model parameters, such as modal amplitudes and decay times.

    \item \textbf{Transient isolation:} After the Z pulse returns to the idle bias, a delay $t_{\mathrm{pre}}$ longer than the short-time ringing separates the RB block from edge-localized transients.

    \item \textbf{RB detection:} After the delay, we apply $m$ random Clifford gates followed by an inversion gate $C_{\mathrm{inv}}$. Residual detuning during this sequence changes the measured RB survival probability.

    \item \textbf{Optimization:} A Nelder--Mead search updates $\boldsymbol{\theta}$ to maximize the mean Clifford fidelity extracted from RB.
\end{enumerate}

The optimized parameters $\boldsymbol{\theta}^\star$ define the calibrated inverse response $\tilde{H}_{\mathrm{LT}}^{-1}(z;\boldsymbol{\theta}^\star)$. Applying this filter reduces the RB-detected effect of residual long-time distortion over the calibrated interval.

\subsection{Cryogenic short-time step response via pre-distortion-assisted Cryoscope}
\label{subsec:supp_cryoscope_shorttime}

After long-time compensation, we use the phase-sensitive qubit probing principle of Cryoscope~\cite{rol2020time} to resolve the remaining ringing near the pulse edges. A time-dependent frequency shift accumulates a Ramsey phase $\phi(\tau)=\int_0^\tau\Delta\omega_q(t)\,\mathrm{d}t$~\cite{blais2021circuit}. Thus, differentiating the measured phase with respect to the evolution time gives the transient frequency response.

Figure~\ref{fig:calib_circuit}(c) and (d) show the preprocessing and measurement sequence:

\begin{enumerate}
    \item \textbf{Pre-distortion preprocessing:} The ideal Z step is processed by $\tilde{H}_{\mathrm{RT}}^{-1}$ and $\tilde{H}_{\mathrm{LT}}^{-1}$, suppressing the long-time components before the short-time measurement.

    \item \textbf{Ramsey sequence:} The preprocessed Z step is applied between an initial $R_x(\pi/2)$ pulse and a final $R_\phi(\pi/2)$ analysis pulse. We scan the Ramsey free-evolution time $\tau$ to measure the accumulated phase at successive times during the pulse.

    \item \textbf{Phase reconstruction:} Measurements with orthogonal analysis phases provide the two quadratures used to reconstruct the accumulated phase $\phi(\tau)$. After phase unwrapping, numerical differentiation yields $\Delta\omega_q(\tau)=\mathrm{d}\phi/\mathrm{d}\tau$, which is converted to the on-chip flux response using the calibrated flux--frequency relation.
\end{enumerate}

The short-time measurement supplies the remaining response information for the FDI-TDF reconstruction in section~\ref{subsec:fdi_tdf}. We reconstruct the compensation impulse response by frequency-domain inversion and fit it with the parallel IIR/FIR model. In this representation, the IIR sections describe the long-time correction, while the FIR branch captures the remaining short-time structure.

\section{Parallel IIR filter coefficient calculation}
\label{app:iir}

The recursive dependence of each IIR output on preceding outputs limits the throughput of a serial implementation. To overcome this constraint, look-ahead block processing expands the recurrence so that a block of $L$ outputs depends only on the preceding $N$ outputs and the current feedforward terms. This appendix derives the coefficient matrices needed for this parallel computation.

\subsection{System definition}

Consider an $N$th-order IIR filter,

\begin{equation}
y[n] = -\sum_{k=1}^N a_k y[n-k] + \sum_{k=0}^{M} b_k x[n-k].
\end{equation}

Define the feedforward term $f[n]$ and write

\begin{equation}\label{eq:origin}
y[n] = -\sum_{k=1}^N a_k y[n-k] + f[n].
\end{equation}

Here, $f[n]=\sum_{k=0}^M b_kx[n-k]$ is the feedforward convolution, $a_k$ is a feedback coefficient, and $b_k$ is a feedforward coefficient.

\subsection{Objective of parallelization}

For $L$-way parallelization, each processing cycle computes

\begin{equation}
\mathbf{y}_{[n:n+L-1]} =
\begin{bmatrix}
y[n] \\
y[n+1] \\
\vdots \\
y[n+L-1]
\end{bmatrix},
\end{equation}
where $n=kL$, $k$ is the processing-cycle index, and $L$ is the parallelization factor. Because $y[n+m]$ depends on the preceding $N$ outputs, we write the block recurrence as

\begin{equation}
\mathbf{y}_{[n:n+L-1]} = \mathit{A}\,\mathbf{y}_{[n-N:n-1]} + \mathit{B}\,\mathbf{f}_{[n:n+L-1]}.
\label{eq:iir_parallel_matrix}
\end{equation}

Here, $\mathbf{y}_{[n-N:n-1]}=[y[n-N],\ldots,y[n-1]]^{\mathrm T}\in\mathbb{R}^{N}$ contains the retained outputs in chronological order, $\mathbf{f}_{[n:n+L-1]}\in\mathbb{R}^{L}$ contains the known feedforward terms, and $\mathit{A}\in\mathbb{R}^{L\times N}$ and $\mathit{B}\in\mathbb{R}^{L\times L}$ are the coefficient matrices.

Let $\alpha_{m,k}$ multiply $y[n-k]$ and $\beta_{m,\ell}$ multiply $f[n+\ell]$ in output $y[n+m]$. For $m=0,1,\ldots,L-1$,

\begin{equation}
\label{eq:target}
y[n+m] = \sum_{k=1}^{N} \alpha_{m,k} y[n-k] + \sum_{\ell=0}^{m} \beta_{m,\ell} f[n+\ell].
\end{equation}

For $m=0$, $\alpha_{0,k}=-a_k$ and $\beta_{0,0}=1$.

\subsection{z-domain derivation of the look-ahead expansion}
Starting from equation~\eqref{eq:origin}, move the feedback terms to the left-hand side and take the $z$-transform:
\begin{equation}
\Bigl(1+\sum_{k=1}^{N} a_k z^{-k}\Bigr) Y(z)=F(z).
\end{equation}

For convenience, define
\begin{subequations}\label{eq:AzYF}
\renewcommand{\theHequation}{\theparentequation.\alph{equation}}
\begin{gather}
A(z) \coloneqq 1+\sum_{k=1}^{N} a_k z^{-k}, \quad (a_0\equiv 1), \label{eq:AzYF:def}\\
A(z)\,Y(z) = F(z). \label{eq:AzYF:main}
\end{gather}
\end{subequations}

To express $y[n+m]$ in terms of retained outputs and known feedforward terms, introduce the look-ahead polynomial of degree at most $m$ in $z^{-1}$
\begin{equation}
Q_m(z)\triangleq 1-\sum_{j=1}^{m}u_j\,z^{-j},
\end{equation}
whose coefficients \(\{u_j\}\) are chosen so that the coefficients of \(z^{-1},\ldots,z^{-m}\) in the product \(Q_m(z)A(z)\) vanish. Expanding
\begin{equation}\label{eq:QmAz}
Q_m(z)A(z)
=\Bigl(1-\sum_{j=1}^{m}u_j z^{-j}\Bigr)\Bigl(\sum_{r=0}^{N} a_r z^{-r}\Bigr).
\end{equation}
With $a_q=0$ for indices outside $0\le q\le N$, equating the coefficients of $z^{-r}$ to zero for $1\le r\le m$ gives the lower-triangular Toeplitz system
\begin{subequations}\label{eq:u-first-terms}
\renewcommand{\theHequation}{\theparentequation.\alph{equation}}
\begin{align}
u_1 &= a_1, \label{eq:u-first-terms:a}\\
u_2 &= a_2 - u_1 a_1, \label{eq:u-first-terms:b}\\
u_3 &= a_3 - u_1 a_2 - u_2 a_1, \label{eq:u-first-terms:c}
\end{align}
\end{subequations}
and, in general,
\begin{equation}
\label{eq:u-rec-general}
u_r = a_r - \sum_{j=1}^{r-1} u_j\, a_{r-j}, \qquad r=2,\ldots,m.
\end{equation}

After the first \(m\) low-order feedback terms have been annihilated, \eqref{eq:QmAz} reduces to

\begin{subequations}\label{eq:QmAz_reduced}
\renewcommand{\theHequation}{\theparentequation.\alph{equation}}
\begin{gather}
Q_m(z)A(z)=1+\sum_{k=1}^{N}\tilde a_{m+k}\, z^{-(m+k)},
\\
\tilde a_{m+k}=a_{m+k}-\sum_{j=1}^{m}u_j\, a_{m+k-j},\label{eq:FA_tildea_single}
\end{gather}
\end{subequations}

Left-multiplying \eqref{eq:AzYF:main} by \(Q_m(z)\) and writing

\begin{equation}\label{eq:q_coeff_single}
Q_m(z)=\sum_{t=0}^{m}c_t^{(m)} z^{-t},
\end{equation}
where
\begin{equation}
    c_0^{(m)}=1, \qquad c_t^{(m)}=-u_t\ (t\ge 1),
\end{equation}
gives
\begin{equation}
\Bigl[1+\sum_{k=1}^{N}\tilde a_{m+k}\, z^{-(m+k)}\Bigr]Y(z)
=\Bigl[\sum_{t=0}^{m}c_t^{(m)} z^{-t}\Bigr]F(z).
\end{equation}

Applying the inverse $z$-transform and evaluating the recurrence at $n+m$ gives
\begin{equation}
y[n+m]+\sum_{k=1}^{N}\tilde a_{m+k}\, y[n-k]
=\sum_{t=0}^{m} c_t^{(m)}\, f[n+m-t].
\end{equation}

Comparing coefficients with the target expansion \eqref{eq:target}, we identify the closed-form look-ahead coefficients:
\begin{widetext}
\begin{subequations}\label{eq:coff}
\renewcommand{\theHequation}{\theparentequation.\alph{equation}}
\begin{gather}
\alpha_{m,k}= -\,\tilde a_{m+k}
= -\,a_{m+k}+\sum_{j=1}^{m}u_j\, a_{m+k-j},\qquad 1\le k\le N,
\\[0.65em]
\beta_{m,\ell}=
\begin{cases}
1,& \ell=m,\\[3pt]
-\,u_{m-\ell},& 0\le \ell<m,\\[3pt]
0,& \ell>m.
\end{cases}
\end{gather}
\end{subequations}
\end{widetext}
The lag index $k$ in equation~\eqref{eq:target} runs in the opposite direction to the chronological column index $j$ of the retained state. The matrices in equation~\eqref{eq:iir_parallel_matrix} are therefore obtained from
\begin{equation}
\mathit{A}_{m+1,j}=\alpha_{m,N-j+1},\qquad
\mathit{B}_{m+1,\ell+1}=\beta_{m,\ell}.
\label{eq:matrix_mapping}
\end{equation}
Here, $m=0,\ldots,L-1$, $j=1,\ldots,N$, and $\ell=0,\ldots,L-1$. In particular, the first row of $\mathit{A}$ is $[-a_N,\ldots,-a_1]$, which reproduces the serial recurrence for $y[n]$ with this state ordering. For the second-order sections and six-sample blocks in figure~\ref{fig:method}(c), $N=2$ and $L=6$, giving a $6\times2$ feedback matrix $\mathit{A}$ and a $6\times6$ feedforward matrix $\mathit{B}$.

%% file: discussion.tex
\section{Discussion and conclusion}

In summary, we have proposed FDI-TDF to reconstruct the compensation impulse response and fit it with a compact IIR/FIR filter for resource-efficient real-time flux distortion compensation. Using look-ahead transformations, we have designed and implemented a parallel FPGA architecture that processes synthesized waveforms at \SI{1.2}{\giga\sample\per\second}. On the OriginQ Wukong processor, the fourth-order, 44-bit IIR configuration achieves a median CZ Pauli fidelity that closely matches the $99.57\%$ software reference.

Numerical analysis and FPGA synthesis establish a quantitative relation between compensation timescale, arithmetic precision, and implementation cost for a sixth-order IIR structure. Hardware resource use grows approximately logarithmically over compensation timescales spanning microseconds to hundreds of milliseconds. Extending coverage from microsecond to hundred-microsecond timescales requires only about $14\%$ more LUT resources and $4\%$ more DSP resources, with a relative arithmetic error below $10^{-4}$. This modest cost of extending temporal coverage supports continuous compensation of long-lived flux responses within limited hardware resources.

Looking ahead, the precision and resource analysis presented here provides a basis for scaling real-time flux compensation to superconducting processors with thousands of physical qubits. Integration with classical feedforward~\cite{caune2026demonstrating} could support calibrated flux control~\cite{sivak2026reinforcement} across measurement-dependent execution paths and repeated quantum error-correction cycles~\cite{eickbusch2025demonstration,besedin2026lattice}. These capabilities could help translate high-fidelity gate control into reliable execution of large-scale dynamic circuits, supporting the development of fault-tolerant superconducting quantum computers.